# Ultra-low magnetization and hysteresis loss in APC Nb₃Sn superconductors

X Xu[1,*], F Wan[1], X Peng[2], and M Sumption[3]

[1] Fermi National Accelerator Laboratory, Batavia, IL 60510, U.S.A

[2] Hyper Tech Research Incorporated, Columbus, OH 43228, U.S.A

[3] The Ohio State University, Columbus, OH 43210, U.S.A

*Author to whom any correspondence should be addressed. E-mail: xxu@fnal.gov

**Abstract**

For the accelerator magnets of the next hadron collider, reducing superconductor persistent-current magnetization is not only important for achieving the desired field quality, but also crucial for its sustainability because the magnetization loss is the major heat load to the magnet cold mass. For conventional Nb₃Sn conductors this requires reduction of effective subelement size ($D_{eff}$). For the restacked-rod-process (RRP®) conductors a physical subelement size ($D_{sub}$) as small as 35 µm (corresponding to a $D_{eff}$ close to 45 µm) can be reached, but at a significant price in $J_c$. Another way to reduce the magnetization is by introducing artificial pinning centers (APC) using the internal oxidation approach. APC conductors outperform conventional Nb₃Sn wires in two aspects: 1) higher $J_c$ at high fields, and 2) much lower $J_c$ and magnetization at low fields (e.g., below 5 T). In this work we explored the fabricability of APC wires with small $D_{sub}$. A 180-stack APC wire was produced and drawn to 0.7- and 0.5-mm diameters with good quality, with $D_{sub}$s of 34 and 24 µm ($D_{eff}$s of 36 and 25 µm), respectively. For the 34-µm-$D_{sub}$ wire, its non-Cu $J_c$ is higher than that of an RRP® wire used for the High-Luminosity Large Hadron

Collider (HL-LHC) project above 13 T (e.g., 36% higher at 4.2 K, 18 T), while its non-Cu magnetization at 1 T, $\Delta M$(1 T), is only 29% of the RRP® wire. Its non-Cu hysteresis loss for a cycle between 1 and 14 T, $Q_h$(1-14 T), is 37% of the RRP® wire. For the 24-µm-$D_{sub}$ wire, its non-Cu $J_c$ surpasses the HL-LHC RRP® wire above 17.5 T, while its $\Delta M$(1 T) and $Q_h$(1-14 T) are only 17% and 23% of the RRP® wire, respectively. Its non-Cu $Q_h$(±3 T) even meets the specification of the International Thermonuclear Experimental Reactor (ITER) project.

## 1. Introduction

Nb$_3$Sn superconductors are being used to produce the interaction-region quadrupole magnets for the High-Luminosity Large Hadron Collider (HL-LHC) project [1,2]. Successful operation of the HL-LHC would confirm the usability of Nb$_3$Sn technology for accelerator magnets in colliders and provide a springboard for future energy-frontier colliders such as the planned Future Circular Collider (FCC)-hh [3]. The baseline target bore field for the FCC-hh dipole magnets was previously 16 T [4] but has been recently reduced to 14 T [5] to increase technical viability. As a result, the specification for the conductor non-Cu $J_c$ has been reduced from 1500 A/mm$^2$ to 1200 A/mm$^2$ at 4.2 K, 16 T (or 705 A/mm$^2$ at 4.2 K, 18 T) [5,6], which can be met by some existing state-of-the-art Nb$_3$Sn conductors – the restacked-rod-process (RRP®) type. RRP® wires have been proven to be reliable conductors with excellent $J_c$ and mechanical properties.

Nevertheless, a much more stringent requirement on conductor magnetization is expected for FCC-hh for a couple of reasons. First, conductor magnetization leads to field errors in the magnet bore. Although magnetic shims can be used to correct some of the field errors, a large magnetization would make sufficient correction challenging, especially if there are flux jumps due to the large magnetization [7]. Second, reducing the magnetization loss is crucial for the sustainability, which is now a critical consideration for any potential future colliders. Magnetization loss is the major heat load to the magnet cold mass and accounts for a significant fraction of the total FCC-hh operational energy consumption [8,9]. In addition, a higher heat load to the cold mass also leads to a requirement for a larger helium inventory, higher helium loss, as well as a higher capital cost for installing cryogenic cooling power [8,9]. The AC losses of a magnet include both persistent-current magnetization loss (i.e., "hysteresis loss") as well as coupling losses, but the hysteresis loss dominates for $Nb_3Sn$ dipole and quadrupole magnets [10] because their typical field ramp rate is low (about 10 mT/s) [5] and the use of Rutherford cables with stainless steel cores significantly suppresses inter-strand coupling [11]. Calculations by Bermudez et al. [10] showed that the hysteresis losses of two-aperture FCC-hh dipole magnets in a full cycle were around 20 kJ/m (depending on the magnet design) if RRP® conductors with an effective subelement size ($D_{eff}$) of 50 µm were used, and that if the $D_{eff}$ was reduced to 20 µm the losses could be reduced to 8-10 kJ/m (which is still above the FCC-hh target losses of 5 kJ/m [3,10]).

It is, however, important to acknowledge the significant challenge to reducing the $D_{eff}$ of RRP® wires to 20 µm. First it should be noted that $D_{eff}$ is different from the physical subelement size ($D_{sub}$). Previous measurements [12-14] showed that for RRP® wires $D_{eff}$ was about 20-30% larger than $D_{sub}$. In fact the $D_{eff}$ values can be calculated from the equation $D_{eff} = (d_o^3 - d_i^3)/(d_o^2 -$ $d_i^2$), where $d_o$ and $d_i$ are the outer and inner diameters of the current-carrying Nb$_3$Sn layer (assumed to be cylindrical) in the subelements [15]. Both RRP® and powder-in-tube (PIT) Nb$_3$Sn wires have large non-superconducting cores in their subelements. However, RRP® wires have much lower unreacted Nb fraction in the subelements than PIT wires (typically about 5% versus 25%, respectively), so for RRP® wires $d_o$ is very close to $D_{sub}$, while for PIT wires $d_o$ is noticeably lower than $D_{sub}$. Using the equation above it can be calculated that for RRP® wires $D_{eff} \approx 1.25 D_{sub}$, agreeing well with the measurement results in [12-14]. Therefore, for RRP® wires the assumed $D_{eff}$s of 50 and 20 µm mentioned above mean $D_{sub}$s around 40 and 16 µm, respectively. So far the smallest $D_{sub}$ reported for RRP® wires is 35 µm [16], which corresponds to a $D_{eff}$ close to 45 µm. However, even using an optimized heat treatment aiming to control the Nausite (a Cu-Nb-Sn phase formed during the heat treatment, which eventually converts to disconnected or coarse-grain Nb$_3$Sn that carries little supercurrent) [17], the RRP® wires with $D_{sub}$ of 35 µm only have non-Cu $J_c$s in the 1100-1300 A/mm$^2$ range at 4.2 K, 15 T [16], lower than or at best on the level of the HL-LHC specification [18], which is significantly lower than the present $J_c$ specification for FCC-hh 14 T dipoles given in [5,6]. Due to the decrease of $J_c$ with decreasing $D_{sub}$, it seems difficult to further reduce the $D_{eff}$ of RRP® wires below 45-50 µm if sufficient $J_c$ is to be maintained.

There are indeed some Nb$_3$Sn conductors that can achieve the low magnetization that the FCC-hh magnets need. Such conductors include the bronze-process type and the single-barrier internal-tin type. Both of them are based on wire designs with very small filaments (below 10 µm in diameter) that are separated by low-Sn bronze after heat treatments (while some filament bridging may occur). They can both achieve whole-wire hysteresis losses (in a +/-3 T cycle) below 500-600 kJ/m$^3$ (corresponding to non-Cu hysteresis losses below 1000-1200 kJ/m$^3$) [19].

This makes these conductors suitable for applications that require super low hysteresis losses, such as the Central Solenoid (CS) and Poloidal Field (PF) coils in tokamak fusion reactors, and other devices requiring fast charging/discharging, such as Superconducting Magnetic Energy Storage (SMES). However, the bronze-process and the single-barrier internal-tin Nb$_3$Sn strands only have moderate $J_c$: their non-Cu $J_c$s are typically only one half or even one third of RRP® wires. For accelerator magnets conductor $J_c$ is crucial because it is a determining factor of the coil size, the conductor amount, and overall, the magnet cost.

On the other hand, an emerging Nb$_3$Sn conductor with artificial pinning centers (APC) based on the internal oxidation technique has demonstrated superior properties compared to state-of-the-art Nb$_3$Sn conductors [20,21]. The APC wires are, to date, the only Nb$_3$Sn conductor that meets the original FCC $J_c$ specification (i.e., with a non-Cu $J_c$ no less than 1500 A/mm$^2$ at 4.2 K, 16 T) [22]. The additional pinning centers in APC wires lead to dramatically improved flux pinning force ($F_p$) and $J_c$ in the Nb$_3$Sn layer [22]. On the other hand, because APC wires have much lower current-carrying Nb$_3$Sn area fraction in the subelements than RRP® wires (35-40% versus ~60%, respectively), we found that the higher non-Cu $J_c$ at high fields in APC wires was not due to higher non-Cu maximum flux pinning force ($F_{p,max}$), but in fact due to a shift of the peak of the $F_p$-$B$ curve to higher fields [23]. Conventional Nb$_3$Sn conductors such as RRP® rely on grain boundaries as flux line pinning centers, and this "surface pinning" leads to $F_p$-$B$ curves peaking at 0.2$B_{irr}$ [24]. In contrast, the oxide particles in APC wires have the right size (mostly 1-10 nm diameter [25-28]) to serve as point pinning centers, and this point pinning leads to $F_p$-$B$ curves peaking at (⅓)$B_{irr}$ [29], which makes the $J_c$-$B$ curves of APC wires much flatter than those of conventional Nb$_3$Sn. With a crossover between the non-Cu $J_c$-$B$ curves of APC and RRP® wires at an intermediate field (e.g., 8-12 T for the conductors in [23]), the flatter $J_c$-$B$ behavior of APC wires results in higher $J_c$ at high fields and simultaneously lower $J_c$ at low fields [23]. For example, magnetization measurements in [23] showed that the ratio of $J_c$(1 T) to $J_c$(16 T) was about 31 for RRP® conductors but only 15 or even lower for APC wires. This means that if RRP® and APC wires reach the same $J_c$ at 16 T, the $J_c$(1 T) of the APC wire would be only half of that of the RRP® wire or even lower [23]. Because the hysteresis loss is mainly dominated by the low-field region, this significantly reduced $J_c$ at low fields would lead to reduced magnetization and hysteresis loss for APC conductors.

Apart from the reduced $J_c$ at low fields, another factor also allows APC conductors to achieve lower magnetization: smaller $D_{eff}$. There are two reasons for this. First, due to the higher unreacted Nb fraction in the subelements of APC wires than in RRP® wires (30-35% versus ~5%, respectively), while $D_{eff} \approx 1.2\text{-}1.3 D_{sub}$ for RRP®, calculations showed that $D_{eff} \approx 1.05\text{-}1.1 D_{sub}$ for APC wires [23]. Second, in RRP® wires each subelement is composed of hundreds of filaments; by contrast, for APC wires based on the PIT design each subelement is composed of only one filament, making the drawability of PIT wires better than RRP® wires. For example, PIT wires with $D_{sub}$ of 25 µm and good quality have been made before [30]. In this work we explored the possibility of fabricating APC wires with very small $D_{sub}$ and measured their magnetizations and hysteresis losses.

## 2. Experimental

*2.1. Samples*

To achieve small $D_{sub}$, an APC wire (Billet number T4276) with 180 filaments was made at Hyper Tech Research Incorporate. To fabricate a filament a mixture of Sn, Cu, and SnO$_2$ powders was filled into a tube with a nominal composition of Nb-7.5wt.%Ta-2wt.%Hf, which was then inserted into a Cu tube to form a subelement. The recipe was similar to the APC wires shown in [31]. After the subelement was cold drawn down and cut into 180 pieces, they were stacked, along with a number of Cu filaments, into a Cu can to form a billet. The non-Cu fraction of this APC wire is about 41%. The billet was cold drawn to diameters of 0.7 mm and 0.5 mm, with corresponding $D_{sub}$s of 34 and 24 µm, respectively. It is worth mentioning that for real applications 0.5-mm-diameter wires are perhaps difficult to use, but we can increase the subelement count to reach the target $D_{sub}$ (e.g., 24-34 µm) at the desired wire diameter (e.g., 0.7-1.0 mm). No wire breakage occurred during wire drawing. Straight segments encapsulated in evacuated quartz tubes were heat treated with a ramp rate of 50°C/h to 400°C, followed by 3°C/h from 400°C to 520°C, then 50°C/h from 520°C to 700°C, and finally kept at 700°C for 25 h and 12 h for the 34-µm-$D_{sub}$ and the 24-µm-$D_{sub}$ APC wires, respectively. The low ramp rate in the intermediate temperature range was employed to ensure sufficient oxygen transfer from the $SnO_2$ in the core to the Nb alloy tube. Scanning electron microscopy (SEM) images of the wires after heat treatments are shown in Figure 1. It is seen that the 34-µm-$D_{sub}$ APC wire has good quality, but the 24-µm-$D_{sub}$ APC wire has a few filaments from which Sn diffused out of the niobium barrier during heat treatment. These filaments are indicated by arrows in Figure 1. An RRP® wire (Billet number 00076) used for the HL-LHC project was also included in this study for comparison, which was heat treated using the recommended schedule of 210°C/48 h + 400°C/48 h + 665°C/75 h with a ramp rate of 25°C/h. More information of this RRP® wire can be found in [22]. It is worth mentioning that this RRP® wire used the "reduced tin" design and does not represent the top $J_c$ levels of RRP® wires.

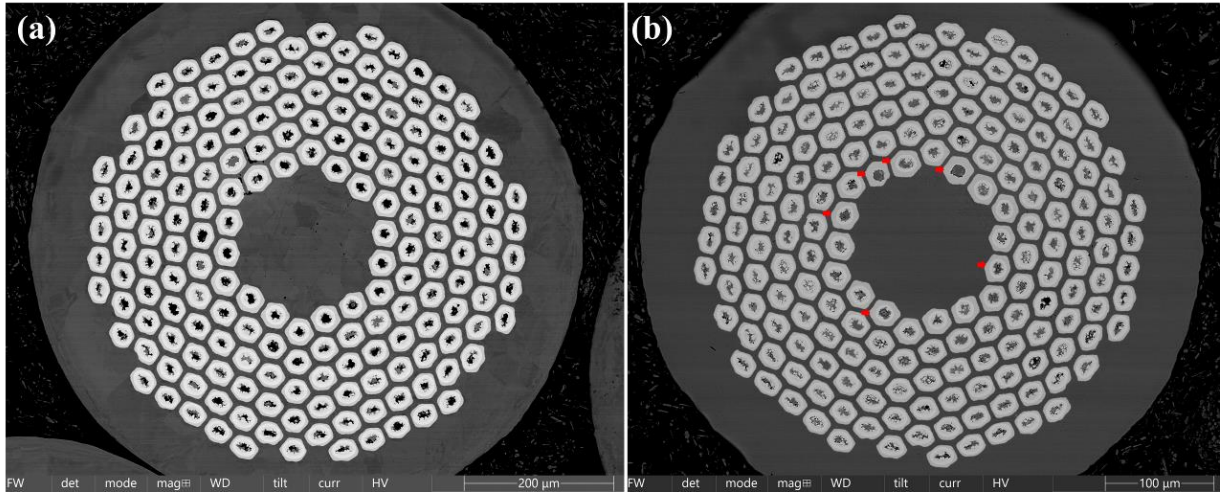

Figure 1. SEM images of (a) the 34-µm-$D_{sub}$ and (b) the 24-µm-$D_{sub}$ APC wires after heat treatments. The filaments from which Sn has diffused into the Cu matrix are indicated by arrows.

## 2.2. Measurements

Voltage versus current (*V-I*) measurements were conducted on straight samples to determine their critical current ($I_c$) values as a function of magnetic field. The RRP® wire and the 34-µm-$D_{sub}$ APC wire were tested in a resistive magnet at the National High Magnetic Field Laboratory (NHMFL) at 4.2 K up to 25 T (while the 24-µm-$D_{sub}$ APC wire missed the scheduled NHMFL testing window), then all of the three wires were also tested at Fermi National Accelerator Laboratory (FNAL) in a 15 T solenoid magnet at 4.2 K. Samples tested at NHMFL were 35 mm in length and at FNAL were 45 mm, both with voltage tap separation of 5 mm. A criterion of 0.1 µV/cm was used to determine the $I_c$ values. The samples for the *V-I* tests were later used for measurements of residual resistivity ratio (RRR) at zero field at FNAL. High RRR values (>80-100) are needed for maintaining the conductor electromagnetic stability. For RRR measurements a current of 1 A was used, and the voltage tap separation was 5 mm. The RRR values were calculated as the ratios of the resistances at room temperature (about 295 K) to those at 20 K: i.e., $R(295\text{ K})/R(20\text{ K})$. It is worth noting that the RRR of Nb$_3$Sn conductors is more commonly defined as $R(293\text{ K})/R(20\text{ K})$. Using tabulated copper resistivity data as a function of temperature (e.g., [32]), it can be calculated that $R(295\text{ K})$ is nearly 1% higher than $R(293\text{ K})$. Therefore, the RRR values reported in this article are nearly 1% higher than those determined based on the $R(293\text{ K})/R(20\text{ K})$ definition.

Magnetization versus field (*M-H*) loops of the samples were measured at 4.2 K using a Vibrating Sample Magnetometer (VSM) associated with a 14 T Physical Properties Measurement System (PPMS) at FNAL. The samples were about 4 mm long, and the magnetic field was perpendicular to the wire length. The magnetic field was ramped with a rate of 12 mT/s using magnetic field sequences of either 0 → -1 → +14 → 0 T or 0 → -3 → +3 → 0 T. The hysteresis loss between two fields was calculated as the area enclosed by the *M-H* loop (i.e., by integrating ∫*M*d*H*) within that range [15]. It is worth mentioning that the *M-H* loop of the RRP® sample measured in this work was slightly different from what we reported in [23], perhaps because of different measurement systems and different segments of samples measured.

## 3. Results

### 3.1. RRR and $J_c$ values

The measured RRR values of the RRP® wire, the 34-µm-$D_{sub}$ and the 24-µm-$D_{sub}$ APC wires were 106, 197, and 87, respectively. The measured non-Cu $J_c$ values of the samples are shown as a function of field in Figure 2. It can be seen that the data measured at NHMFL and at FNAL are in agreement. Fittings were made to the $J_c$-*B* data using the general scaling relation: $J_c$ =

$C·(B/B_{c2}^*)^p·(1-B/B_{c2}^*)^q/B$, where $C$, $B_{c2}^*$, $p$ and $q$ are fitting parameters [29]. For the RRP® wire, $p$ = 0.5 and $q$ = 2 were used. It can be seen that the fitted curves match the data points well. For the 24-µm-$D_{sub}$ APC wire, despite the sufficient number of data points to perform the fit, due to the limited field range for the data its fitted curve should be interpreted with caution, particularly in the regions far from the data points. The obtained fitted parameters are summarized in Table 1.

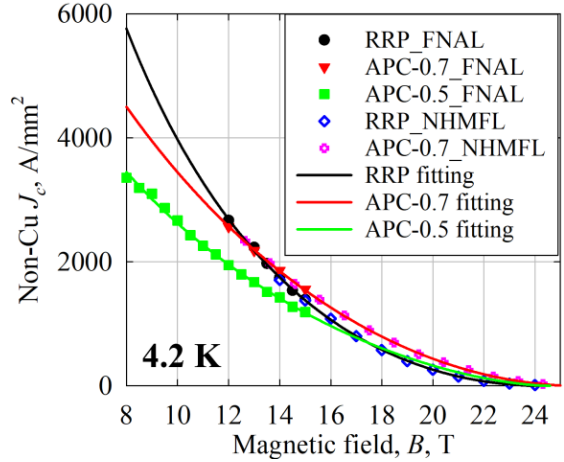

Figure 2. Non-Cu $J_c(B)$ data of the RRP® and the APC wires at 4.2 K.

Table 1. Summary of the obtained fitted parameters for the samples.

| The fitted parameters | $C$, A*T/mm² | $B_{c2}^*$, T | $p$ | $q$ |
| --- | --- | --- | --- | --- |
| The RRP® wire for HL-LHC | 178185 | 24.4 | 0.5 | 2 |
| The 34-µm-$D_{sub}$ APC wire | 194425 | 25.7 | 0.83 | 1.92 |
| The 24-µm-$D_{sub}$ APC wire | 115631 | 24.6 | 0.71 | 1.63 |

*3.2. Magnetization*

The *M-H* loops of the samples in the 0-14 T range are shown in Figure 3. In Figure 3(a) the magnetizations are normalized to the whole-wire volumes as whole-strand magnetization is the most commonly used in literature. But because the APC wires here have lower non-Cu fraction than the RRP® wire, to make a fair comparison the non-Cu *M-H* loops are shown in Figure 3(b).

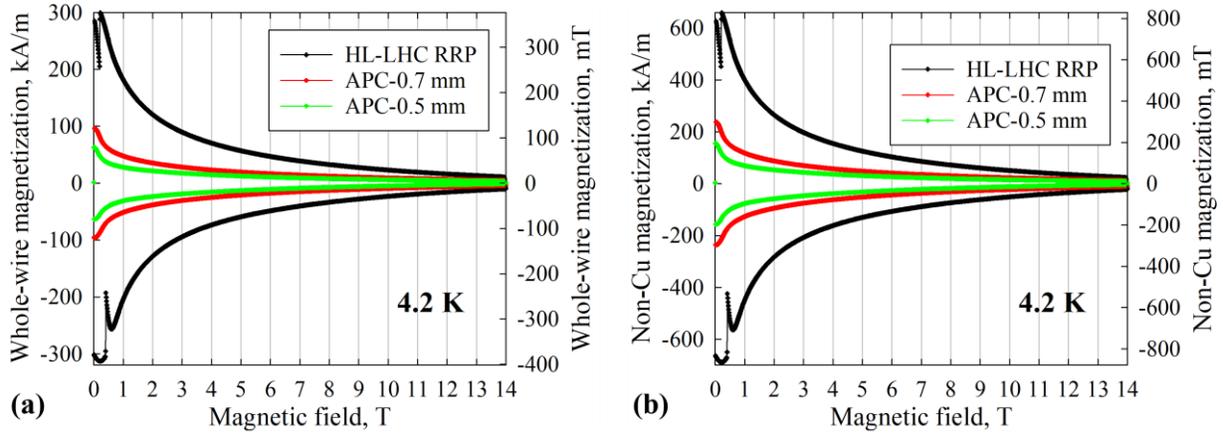

Figure 3. (a) Whole-wire and (b) non-Cu *M-H* loops of the RRP® and the APC wires.

In order to see how the magnetization and hysteresis loss of the 24-µm-$D_{sub}$ APC wire compare to the bronze-process and the single-barrier internal-tin strands, its *M-H* loop in the +/- 3 T range, normalized to both the whole-wire and non-Cu volumes, is shown in Figure 4.

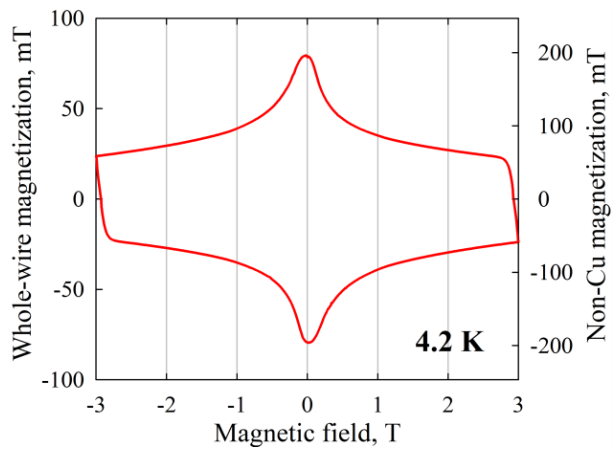

Figure 4. Whole-wire and non-Cu *M-H* loop (+/-3 T) of the 24-µm-$D_{sub}$ APC wire. The non-Cu hysteresis loss in a +/-3 T cycle is calculated to be 847 kJ/m$^3$.

## 4. Discussions

The RRR values of the 34-µm-$D_{sub}$ and the 24-µm-$D_{sub}$ APC wires (197 and 87, respectively) indicate that the wire quality was good for a $D_{sub}$ of 34 µm but degraded as the $D_{sub}$ was reduced to 24 µm, although a RRR of 87 is still respectable. This can also be seen from the SEM images in Figure 1: in the 24-µm-$D_{sub}$ APC wire Sn diffused out from a few filaments into the Cu matrix during heat treatment. These filaments are in the innermost layer of the wire – we also noticed the same phenomenon in other APC wires: the innermost-layer filaments tend to have a lower quality than those in the outer layers. A solution to this issue for future small-$D_{sub}$ APC wires is to use more conservative recipes for the innermost-layer filaments.

In Figure 2 the non-Cu $J_c$-$B$ curves of the 34-µm-$D_{sub}$ APC wire and the RRP® wire intersect around 13 T. The APC wire has higher non-Cu $J_c$ above 13 T, and in fact, the higher the field is, the more advantage the APC wire has. For example, its non-Cu $J_c$ is only 12% higher than the RRP® wire at 15 T, but at 18 T it is 36% higher. Of course, due to the lower non-Cu fraction in the APC wire (41%) compared to the RRP® wire (45.5%), the advantage of the APC wire at 18 T is less in terms of engineering critical current density $J_e$ (i.e., $I_c$ divided by the wire's entire cross-sectional area). The $J_e$ values of the APC wire and the RRP® wire at 4.2 K, 18 T are 325 and 265 A/mm$^2$, respectively. It is worth mentioning, however, that the non-Cu fraction can be adjusted without much technical difficulty by using Cu tubes with a different thickness and by changing the number of Cu filaments in a billet. The non-Cu $J_c$ of this APC wire is lower than the APC wires reported in our previous papers [22,31] due to the much smaller $D_{sub}$, indicating that APC wires, just like the conventional PIT and RRP® wires [7], also show a decrease in $J_c$ with decreasing $D_{sub}$. However, its non-Cu $J_c$ (about 800 A/mm² at 4.2 K, 18 T) exceeds the $J_c$ specification for the FCC-hh 14 T dipoles given in [6] (for comparison, that of the HL-LHC RRP® wire used in this study is noticeably below this specification). For the 24-µm-$D_{sub}$ APC wire, although its non-Cu $J_c$ is lower than that of the 34-µm-$D_{sub}$ wire, it is expected to surpass the RRP® wire above 17.5 T based on the fitted $J_c$-B curves in Figure 2. Its fitted non-Cu $J_c$ at 4.2 K, 18 T is about 14% below the specification for FCC-hh 14 T dipoles [6].

Here it is worth discussing why $J_c$ at 18 T is a critical parameter for the FCC-hh dipoles with a target operating field of 14 T in bore (for which the peak field on conductors, $B_p$, is typically slightly higher – e.g., 14.3-14.6 T, depending on the magnet design [5]). The baseline design for the FCC-hh 14 T dipoles assumes a 20% margin along the load line (which was chosen based on the magnets of previous machines) [5,6], corresponding to a short sample limit field ($B_{ss}$) around 18 T for the peak field. It is the $J_c$ at $B_{ss}$ that determines the operating current density ($J_{op}$): $J_{op}$= 0.8*$J_c(B_{ss})$. Of course, the $J_c$ at the $B_p$ is also important because the current margin (1 − $J_{op}/J_c$ at the $B_p$) is needed for current sharing. However, it is the load line margin, not the current margin, that is related to the temperature margin (and thus the energy margin), which are critical for magnet operation against perturbations [33]. For this reason, the $J_c$ at $B_{ss}$ (i.e., ~18 T) is a more relevant parameter than the $J_c$ at $B_p$ (i.e., ~14.5 T) for FCC-hh 14 T dipoles. Of course, in magnet operations stress also has an important influence on the conductor critical surface and should be considered.

In Figure 2 at low fields the APC wires have reduced non-Cu $J_c$ relative to the RRP® wire, and the lower the field is, the more significant the $J_c$ reduction is. This reduction of $J_c$ at low fields leads to a reduction of magnetization. The heights of the *M-H* loops (Figure 3) at 1 T, $\Delta M$(1 T), of the three wires are shown in Table 2. Assuming the injection field of the planned FCC-hh is 1 T [7] and the operational field is 14 T, the hysteresis losses between 1 and 14 T were calculated and also listed in Table 2. To make the comparison fair the non-Cu values are shown here. In addition, the $D_{sub}$ and $D_{eff}$ values for these conductors are also summarized in Table 2.

Table 2. Summary of the magnetization, hysteresis loss, $D_{sub}$, and $D_{eff}$ values of the wires.

|  | RRP® for HL-LHC | 34-µm-$D_{sub}$ APC | 24-µm-$D_{sub}$ APC |
|---|---|---|---|
| Non-Cu $\Delta M$(1 T), mT | 1065 | 309 | 183 |
| Non-Cu hysteresis loss (1 → 14 → 1 T), kJ/m$^3$ | 2908 | 1091 | 664 |
| $D_{sub}$, µm | 55 | 34 | 24 |
| $D_{eff}$, µm | 69 | 36 | 25 |

The non-Cu $\Delta M$(1 T) of the 34-µm-$D_{sub}$ APC wire is only 29% of the RRP® wire. Due to the lower magnetization, it also has lower hysteresis loss: only 37% of the RRP® wire in a 1-14 T cycle. The 24-µm-$D_{sub}$ APC wire has even lower magnetization: its non-Cu $\Delta M$(1 T) and hysteresis loss in a 1-14 T cycle are only 17% and 23% of the RRP® wire, respectively. In fact, its non-Cu hysteresis loss in a +/-3 T cycle calculated from Figure 4, about 847 kJ/m$^3$, is comparable to some bronze-process and single-barrier internal-tin Nb$_3$Sn conductors [19], and also meets the specification for the International Thermonuclear Experimental Reactor (ITER) project, which requires the non-Cu hysteresis loss in a +/-3 T cycle to be below 1000 kJ/m$^3$ [19]. On the other hand, its non-Cu $J_c$ is significantly higher than bronze-process and single-barrier internal-tin strands (especially at high fields), making APC wires suitable for high-field applications that require low hysteresis losses.

The reduction of conductor hysteresis loss discussed in this article has significant implications for future hadron colliders. As shown in [34], for 14 T dipole magnets operated at 1.9 K (the present baseline for FCC-hh), the magnet losses, if assumed to be 10 kJ/m, would contribute an energy consumption similar to that of synchrotron radiation. However, as noted in the "*Introduction*" section, due to the difficulty in reducing the $D_{eff}$ of RRP® wires below 45-50 µm while maintaining adequate $J_c$, a more realistic target for the magnet losses using RRP® conductors is 20 kJ/m – approximately twice the energy consumption from the synchrotron radiation. By contrast, this work demonstrates that APC conductors can reduce the hysteresis loss by approximately two-thirds. If these APC conductors could be employed for building the 1.9 K, 14 T dipoles, the total energy consumption (i.e., synchrotron radiation plus the magnet losses) could be reduced by nearly half. Efforts are currently underway to scale up the production of such low-$D_{eff}$ APC conductors. Finally, it is worth noting that heat treatment temperature also has a considerable influence on the $F_p$-$B$ curve peak shift in APC wires [27,35]. Further studies are still needed to optimize the heat treatments of APC wires [34,36]. In addition, as discussed earlier, stress also influences the conductor critical surface, so the stress tolerance of APC conductors needs to be characterized as well.

## 5. Conclusions

In this work we fabricated a 180-stack APC wire and drew it to 0.7 and 0.5 mm diameters (with $D_{sub}$s of 34 and 24 µm, respectively) to reduce the magnetization and hysteresis loss. The 34-µm-$D_{sub}$ APC wire had good quality with RRR of 197. It has a flatter $J_c$-$B$ curve than an HL-LHC RRP® wire, with the crossover around 13 T. Above 13 T it has higher $J_c$ than the RRP® wire (about 12% and 36% higher at 15 T and 18 T, respectively). Meanwhile, the APC wires have much lower magnetizations and hysteresis losses. The non-Cu $\Delta M$(1 T) and the non-Cu hysteresis loss in a 1-14 T cycle of the 34-µm-$D_{sub}$ APC wire are 29% and 37% of the RRP® wire, respectively, while those of the 24-µm-$D_{sub}$ APC wire are only 17% and 23% of the RRP® wire, respectively. With a non-Cu hysteresis loss of only 847 kJ/m$^3$ in a +/-3 T cycle, the 24-µm-$D_{sub}$ APC wire even meets the specification for the ITER conductors, but has a significantly higher non-Cu $J_c$. There are three reasons for the significant reduction of low-field magnetization in the APC wires: (1) lower $J_c$ at low fields due to the flatter $J_c$-$B$ curves, (2) smaller $D_{sub}$, (3) $D_{eff}$–$D_{sub}$ relation (for RRP® wires $D_{eff} \approx 1.25 D_{sub}$, but for the APC wires $D_{eff}$ is quite similar to $D_{sub}$ due to the higher unreacted Nb fraction in the subelements). Because it is difficult to reduce the $D_{sub}$ of RRP® wires below 35 µm (corresponding to a $D_{eff}$ close to 45 µm) given that RRP® wires with $D_{sub}$ of 35 µm already have a significantly reduced $J_c$, the hysteresis losses of two-aperture FCC-hh dipole magnets built with RRP® conductors will most likely be well above 10 kJ/m for a full cycle. On the other hand, the APC wires are promising to achieve the targets of magnetization and hysteresis loss (e.g., 5 kJ/m) for FCC-hh dipole magnets, which is important for its sustainability.

**Acknowledgements**

This work was supported by the US Department of Energy through an Early Career Program Award. This work was produced by Fermi Forward Discovery Group, LLC under Contract No. 89243024CSC000002 with the U.S. Department of Energy, Office of Science, Office of High Energy Physics. Publisher acknowledges the U.S. Government license to provide public access under the DOE Public Access Plan. A portion of this work was performed at NHMFL, which is supported by National Science Foundation Cooperative Agreement No. DMR-1644779 and the State of Florida.

**References**

[1]. Bermudez S I *et al.* 2023 Status of the MQXFB Nb$_3$Sn quadrupoles for the HL-LHC *IEEE Trans. Appl. Supercond.* **33** 4001209

[2]. Ambrosio G *et al.* 2023 Challenges and lessons learned from fabrication, testing, and analysis of eight MQXFA low beta quadrupole magnets for HL-LHC *IEEE Trans. Appl. Supercond.* **33** 4003508

[3]. Benedikt M *et al.* 2019 FCC-hh: The Hadron Collider *Eur. Phys. J. Special Topics* **228** 755–1107

[4]. Tommasini D and Toral F 2016 Overview of magnet design options *EuroCirCol-P1-WP5 report* 1-16

[5]. Todesco E 2025 Status and Perspectives of High Field Magnets for Particle Accelerators *IEEE Trans. Appl. Supercond.* **35** 4003914

[6]. Todesco E 2025 Tentative guidelines for magnet design *HFM Forum* https://indico.cern.ch/event/1501797/

[7]. Ballarino A and Bottura L 2015 Targets for R&D on Nb$_3$Sn Conductor for High Energy Physics *IEEE Trans. Appl. Supercond.* **25** 6000906

[8]. Delprat L 2025 FCC-hh cryogenics update for 1.9 K and 4.5 K options *FCC Week 2025* Vienna Austria

[9]. De Sousa P T C B and Cidoncha X G 2025 Cryogenic baseline and advancements on the 4.5 K option for FCC-hh using Nb$_3$Sn magnets *FCC Week 2025* Vienna Austria

[10]. Bermudez S I 2024 Hysteresis losses in EuroCirCol: model and measurements *HFM Forum* https://indico.cern.ch/event/1449701/

[11]. Collings E W, Sumption M D, Susner M A, Dietderich D R, Barzi E, Zlobin A V, Ilyin Y and Nijhuis A 2008 Influence of a Stainless Steel Core on Coupling Loss, Interstrand Contact Resistance, and Magnetization of an Nb$_3$Sn Rutherford Cable *IEEE Trans. Appl. Supercond.* **18** 1301-4

[12]. Ghosh A K, Cooley L D, Moodenbaugh A R, Parrell J A, Field M B, Zhang Y and Hong S 2005 Magnetization studies of high $J_c$ Nb$_3$Sn strands *IEEE Trans. Appl. Supercond.* **15** 3494-7

[13]. Parrell J A, Zhang Y, Field M B and Hong S 2007 Development of internal tin Nb$_3$Sn conductor for fusion and particle accelerator applications *IEEE Trans. Appl. Supercond.* **17** 2560-3

[14]. Turrioni D, Barzi E, Bossert M, Kashikhin V V, Kikuchi A, Yamada R and Zlobin A V 2007 Study of Effects of Deformation in Nb$_3$Sn Multifilamentary Strands *IEEE Trans. Appl. Supercond.* **17** 2710-3

[15]. Sumption M D, Peng X, Lee E, Wu X and Collings E W 2004 Analysis of magnetization, AC loss, and $d_{eff}$ for various internal-Sn based $Nb_3Sn$ multifilamentary strands with and without subelement splitting *Cryogenics.* **44** 711-25

[16]. Barzi E, Turrioni D, Ivanyushenkov Y, Kasa M, Kesgin I, and Zlobin A V 2021 Heat Treatment Studies of $Nb_3Sn$ Wires for Superconducting Planar Undulators *IEEE Trans. Appl. Supercond.* **30** 6001005

[17]. Sanabria C, Lee P J, Larbalestier D C 2016 The vital role of a well-developed Sn-Nb-Cu membrane for high Jc RRP® $Nb_3Sn$ wires, *IEEE Trans. Appl. Supercond.* 1-2

[18]. Cooley L D, Ghosh A K, Dietderich D R and Pong I 2017 Conductor Specification and Validation for High-Luminosity LHC Quadrupole Magnets *IEEE Trans. Appl. Supercond.* **27** 6000505

[19]. Vostner A et al 2017 Statistical analysis of the $Nb_3Sn$ strand production for the ITER toroidal field coils *Supercond. Sci. Technol.* **30** 045004

[20]. Xu X, Sumption M D and Peng X 2015 Internally Oxidized $Nb_3Sn$ Superconductor with Very Fine Grain Size and High Critical Current Density *Adv. Mater.* **27** 1346-50

[21]. Buta F *et al* 2021 Very high upper critical fields and enhanced critical current densities in $Nb_3Sn$ superconductors based on Nb–Ta–Zr alloys and internal oxidation *J. Phys. Mater.* **4** 025003

[22]. Xu X, Peng X, Lee J, Rochester J and Sumption M D 2020 High Critical Current Density in Internally-oxidized $Nb_3Sn$ Superconductors and its Origin *Scr. Mater.* **186** 317-320

[23]. Xu X, Sumption M D, Wan F, Peng X, Rochester J and Choi E S 2023 Significant reduction in the low-field magnetization of $Nb_3Sn$ superconducting strands using the internal oxidation APC approach *Supercond. Sci. Technol.* **36** 085008

[24]. Kramer E J 1973 Scaling laws for flux pinning in hard superconductors *J. Appl. Phys.* **44** 1360

[25]. Rochester J, Ortino M, Xu X, Peng X and Sumption M 2021 The roles of grain boundary refinement and nano-precipitates in flux pinning of APC $Nb_3Sn$ *IEEE Trans. Appl. Supercond.* **31** 8000205

[26]. Ortino M, Pfeiffer S, Baumgartner T, Sumption M, Bernardi J and Xu X 2021 Evolution of the superconducting properties from binary to ternary APC-$Nb_3Sn$ wires *Supercond. Sci. Technol.* **34** 035028

[27]. Xu X, Peng X, Rochester J, Sumption M D, Lee J, Ortiz G A C and Hwang J 2021 The strong influence of Ti, Zr, Hf solutes and their oxidation on microstructure and performance of $Nb_3Sn$ superconductors *J. Alloys Compd.* **857** 158270

[28]. Lee J, Mao Z, Isheim D, Seidman D N, and Xu X 2024 Unveiling the nucleation and growth of Zr oxide precipitates in internally oxidized $Nb_3Sn$ superconductors *J. Alloy. Compd.* **1005** 176044

[29]. Hughes D 1974 Flux pinning mechanisms in type II superconductors *Phil. Mag.* **30** 293-305

[30]. Lindenhovius J L H, Hornsveld E M, den Ouden A, Wessel W A J and ten Kate H H J 2000 Powder-in-Tube (PIT) Nb$_3$Sn conductors for high-field magnets *IEEE Trans. Appl. Supercond.* **10** 975-8

[31]. Xu X, Peng X, Wan F, Rochester J, Bradford G, Jaroszynski J and Sumption M 2023 APC Nb$_3$Sn superconductors based on internal oxidation of Nb-Ta-Hf alloys *Supercond. Sci. Technol.* **36** 035012

[32]. Matula R A 1979 Electrical resistivity of copper, gold, palladium, and silver *J. Phys. Chem. Ref. Data* **8** 1147–1298

[33]. Todesco E and Ferracin P 2012 Limits to high field magnets for particle accelerators *IEEE Trans. Appl. Supercond.* **12** 4003106

[34]. Todesco E 2025 HFM Targets and requirements for FCC hh magnets *2nd High Temperature Superconductor Accelerator Technology (HiTAT-2) Workshop* https://indico.cern.ch/event/1455509/contributions/6586227/

[35]. Lonardo F et al. 2024 Influence of the Heat Treatment on the Layer $J_c$ of Internal-Sn Nb$_3$Sn Wires with Internally Oxidized Nanoparticles *IEEE Trans. Appl. Supercond.* **34** 6000305

[36]. Wan F, Xu X, Peng X and Sumption M D 2025 Phase Evolution and Area Fractions of Coarse-Grain and Fine-Grain A15 in APC Nb$_3$Sn Superconductors *IEEE Trans. Appl. Supercond.* **35** 6000205